\documentclass{phb-proc4-auth}

\usepackage{graphicx}
\usepackage{amssymb}
\usepackage{amsmath}

\newcommand{\Journal}[4]{#1 \textbf{#2}, #3 (#4)}
\newcommand{\PRev}{Phys. Rev.}
\newcommand{\JPSJ}{J. Phys. Soc. Jpn.}
\newcommand{\be}{\begin{equation}}
\newcommand{\ee}{\end{equation}}
\newcommand{\nn}{\nonumber}

\newcommand{\pr}{PrOs$_4$Sb$_{12}$}
\newcommand{\mb}[1]{\mathbf{#1}}

\begin{document}
\begin{frontmatter}


\journal{SCES '04}


\title{
Landau theory of superconducting phases in $T_h$ crystals}

%
%
%
%
%
%

\author{S. H. Curnoe\corauthref{1}}
and
\author{I. A. Sergienko\thanksref{NSERC}}
\address{
Department of Physics \& Physical Oceanography, Memorial University of Newfoundland, St. John's, NL, A1B 3X7, Canada}

%
 

%
%
%
%


%
%
%
%
\thanks[NSERC]{This work was supported by NSERC of Canada.}
\corauth[1]{Email:curnoe@physics.mun.ca}


\begin{abstract}

We consider the Landau theory of phase transitions for multiple superconducting phases in $T_h$ crystals.  All possible phase transition sequences involving a single superconducting order parameter are found.  These results may be applicable to \pr.  

\end{abstract}

%
%



\end{frontmatter}

%
%
%
%
%

Perhaps the greatest descriptive power and beauty of the Landau
theory of phase transitions is revealed when successive phase transitions
are described by a single, multi-component order parameter (OP).
There are many well-known examples of this involving
structural OP's, especially among ferroelectric compounds.  But
although multi-component OP's are the hallmark of most
unconventional superconductors, so far there are no confirmed
examples of multiple superconducting (SC) phases described by a single 
OP.  PrOs$_4$Sb$_{12}$, which displays two SC phases
with different symmetries, may be the first.

The application of Landau theory to multiple SC phases described by a single
OP 
was studied previously~\cite{Gufan1995} for various crystal symmetries, 
and applied to 
UPt$_3$.  
This material displays multiple superconducting phase transitions due to
a two-dimensional SC order parameter, but the degeneracy of
the order parameter is lifted by
antiferromagnetic ordering.
In \pr, superconductivity occurs at $T_{c1} = 1.85$ K;
an additional phase transition is observed 
at $T_{c2}=1.75$ K \cite{Vollmer2003,Tayama03}.
Thermal conductivity measurements 
revealed the presence of nodes and a lowering of the symmetry of the gap
function from four-fold  to two-fold  at
$T_{c2}$\cite{Izawa2003},  which is highly suggestive of a symmetry lowering
phase transition involving the SC order parameter.
In this article, we describe the phenomenological
Landau theory of
multiple SC phases in $T_h$ crystals, which may be applied to \pr.

Especially useful when the microscopic origin of superconductivity is
unknown, the phenomenological approach is based entirely on symmetry.
Order parameters are classified according to representations of the
crystallographic point group, in terms of which the Landau potential
is expanded.  In principle, all possible phases, their symmetries
and phase diagrams, may be obtained in 
this way, as well as the positions and types 
of nodes of the corresponding SC gap functions. 

The point group $T_h$ has one-dimensional (1D), two-dimensional (2D) and
three-dimensional (3D)  representations, called $A$, $E$ and
$T$ respectively, which may be even (for singlet
pairing) or odd (triplet pairing).
The phase diagram for the 1D representation $A$ involves only one
SC phase, 
therefore we do not consider it any further.
For the 
representations $E$ and $T$, we restrict our attention to phases connected by
second order phase transitions, as suggested in at least
one experiment \cite{Vollmer2003}.  Together
with the restriction to a single OP, this sets
strong constraints on allowed phase transition sequences.

The SC gap function is a $2 \times 2$ matrix, 
$\widehat \Delta(\mathbf{k})=i\widehat\sigma_y \psi(\mathbf{k})$ for singlet pairing and 
$\widehat \Delta(\mathbf{k})=i(\mathbf{d}(\mathbf{k})\boldsymbol{\widehat\sigma})\widehat
\sigma_y$ for triplet pairing, where $\boldsymbol{\widehat\sigma}=(\widehat\sigma_x, \widehat
\sigma_y, \widehat\sigma_z)$ are Pauli matrices, $\psi(\mathbf{k})$ is an
even scalar function and $\mathbf{d}(\mathbf{k})$ is
an odd pseudovector function.
The gap in the quasiparticle energy spectrum in the singlet SC state is
given by $\Delta(\mb k) = |\psi(\mb k)|$, while in the triplet state it is
$\Delta_\pm(\mb k) = [|\mb d(\mb k)|^2 \pm |\mb d(\mb k) \times \mb d^*(\mb k)|]^{1/2}$.
The functions $\psi(\mathbf{k})$ and $\mathbf{d}(\mathbf{k})$
are expressed in terms of the components of the OP
$\eta_i$ as
\be
\psi(\mathbf{k})=\sum_i \eta_i \psi_i(\mathbf{k}), \qquad
\mathbf{d}(\mathbf{k})=\sum_i \eta_i \mathbf{d}_i(\mathbf{k}),
\label{gap}
\ee
where $\psi_i(\mathbf{k})$ and $\mathbf{d}_i(\mathbf{k})$ are the basis functions for
the even and odd  representations of the point
group, respectively \cite{Sigrist1991}.

The transformation properties of the OP are derived from the basis functions
$\psi_i$ and $\mathbf d_i$, which may be arbitrarily chosen to span the
representation.   For the 2D representation $E$, we choose  basis functions that are complex
conjugate \cite{serg}.
The Landau potential expanded up to sixth order in the 2D OP 
is
\begin{eqnarray}\label{energy_Eg}
F&=&\alpha(|\eta_1|^2+|\eta_2|^2) + \beta_1 (|\eta_1|^4+|\eta_2|^4) +
2\beta_2 |\eta_1|^2|\eta_2|^2\nn\\
&& + \gamma_1 (\eta_1^3\eta_2^{*3}+\eta_2^3\eta_1^{*3})
+ \gamma_2 i(\eta_1^3\eta_2^{*3}-\eta_2^3\eta_1^{*3}),
\label{F_Eg}
\end{eqnarray}
where $\alpha$, $\beta_1$, $\beta_2$, $\gamma_1$, and $\gamma_2$ are phenomenological parameters.
The Landau potential is invariant under $ K \times U\times G$,
where $K$ is time reversal, $U$ are
$U(1)$  gauge transformations and $G$ is the space group.
Eq.~(\ref{F_Eg}) describes second order phase transitions between the normal
state and the states $(1,0)$ and $(\phi_1,\phi_2)$, as shown
 in Fig. 1a).  The components of the state $(\phi_1,\phi_2)$ have equal
magnitudes and unfixed relative phases.  The sixth order expansion is
insufficient to describe the additional second order phase transition
to the state $(\eta_1,\eta_2)$, which has neither fixed relative magnitudes or
phases.  

To obtain the third phase $(\eta_1,\eta_2)$ 
it is necessary to consider additional
higher order terms in the Landau potential.  However, such an approach is rarely
followed in practice.    
Instead, 
an {\em effective} Landau potential of an {\em effective} OP
is used to analyse phase transitions
between the various SC phases whose symmetries
satisfy group-subgroup relations.  Second order phase transitions are
indicated by the absence  of third order terms of the effective
OP.   In our case, $(\eta_1,\eta_2)$ has symmetry $D_2$, which is a subgroup
of the symmetry group of $(\phi_1,\phi_2)$, $D_2\times K$.
Therefore the effective OP is 1D and breaks time
reversal symmetry, hence odd order terms in the effective Landau potential are
prohibited.  It follows that the transition is second order.   The symmetry of
$(1,0)$ is the group $T(D_2)$ (see Ref. \cite{serg}), which is a supergroup 
of $D_2$.  The transition $(1,0)\rightarrow(\eta_1,\eta_2)$ is described
by an effective 1D OP which is just $\eta_2$.  It is clear
from Eq.~(\ref{F_Eg}) that there will be third order terms in the
effective Landau potential of the form $\eta_2^3\pm \eta_2^{*3}$.  
Therefore the transition  $(1,0)\rightarrow(\eta_1,\eta_2)$ is first order.  

\begin{figure}
\centering
\includegraphics[width=2.2in]{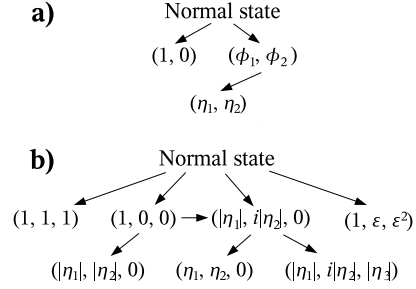}

\caption{
Second order phase transition sequences for a) the 2D order parameter $E$ and b) the 3D order parameter $T$.  This diagram does not show phases or transitions beyond two second order transitions from the normal state.  Note that $\varepsilon = \exp\frac{i2\pi}{3}$ and $(1,\varepsilon,\varepsilon^2)$ has components with equal magnitude and phases fixed by $\varepsilon$.
}
\end{figure}  

For the 3D OP, real basis functions which transform under rotation
as $(x,y,z)$  are
chosen for $\psi_i$ and $\mathbf d_i$ \cite{serg}.
The free energy contains second and fourth order terms of the form\\
$
I_1 = |\eta_1|^2 + |\eta_2|^2 + |\eta_3|^2$, $I_1^2$, $I_2 = |\eta_1|^4+|\eta_2|^4
+|\eta_3|^4$, 
$
I_{3} = \eta_1^2\eta_2^{*2}+\eta_2^2\eta_3^{*2}+
\eta_3^2\eta_1^{*2} + {\mbox c.c.}
$,\\
and sixth order terms of the form\\
$
I_1^3$, $I_1I_2$, $I_1I_3$, $I_4 = |\eta_1|^6+|\eta_2|^6+|\eta_3|^6$, $I_5 =
|\eta_1|^4(|\eta_2|^2-|\eta_3|^2)+|\eta_2|^4(|\eta_3|^2-|\eta_1|^2)+
|\eta_3|^4(|\eta_1|^2-|\eta_2|^2)$,
$I_{6,7} = [\eta_1^4\eta_2^{*2}+\eta_2^4\eta_3^{*2}+\eta_3^4\eta_1^{*2}
   \pm(\eta_2^4\eta_1^{*2}+\eta_3^4\eta_2^{*2}+\eta_1^4\eta_3^{*2})]
+\mbox{c.c.}
$
These terms are sufficient to describe second order phase transitions 
between the normal state and the upper four SC states shown in Fig. 1b).  
As in the case of the 2D OP, higher order terms are needed to describe the 
additional phase transitions shown in Fig. 1b).  
Among the various SC phases satisfying group-subgroup relations, it is
straightforward to determine which of them are connected by second order phase
transitions, given their symmetry groups (see Ref. \cite{serg}).
In all cases, the subgroup contains half the number of 
elements as the group, which means that the transition can be described 
by a 1D effective order parameter which has character $-1$ for half of the
group elements.  Third order terms in the effective potential are therefore
prohibited, and the transitions are second order.
 
Properties of the gap function, especially nodes, may be determined by
substituting the OP for each phase into Eq. (\ref{gap}).  
For \pr, 
this analysis fails to explain the four-fold to two-fold
symmetry lowering of the gap function observed in Ref. \cite{Izawa2003}.
This suggests that the transition at $T_{c2}$ is due either to a second
SC OP, or else to an OP that is of a different origin altogether.

\end{document}